\documentclass[a4paper]{jpconf}
\usepackage{graphicx}
\usepackage{amsfonts}  
\usepackage{amsmath}
\usepackage{amssymb} 

\newcommand{\be}{\begin{equation}}  
\newcommand{\ee}{\end{equation}}
\newcommand{\beq}{\begin{eqnarray}}  
\newcommand{\eeq}{\end{eqnarray}}  
\newcommand{\hH}{\hat{H}}  
\newcommand{\hN}{\hat{N}} 
  
\begin{document}  
      
\def\bbe{\mbox{\boldmath $e$}}  
\def\bbf{\mbox{\boldmath $f$}}      
\def\bg{\mbox{\boldmath $g$}}  
\def\bh{\mbox{\boldmath $h$}}  
\def\bj{\mbox{\boldmath $j$}}  
\def\bq{\mbox{\boldmath $q$}}  
\def\bp{\mbox{\boldmath $p$}}  
\def\br{\mbox{\boldmath $r$}}      
  
\def\bone{\mbox{\boldmath $1$}}      
  
\def\dr{{\rm d}}  
  
\def\tb{\bar{t}}  
\def\zb{\bar{z}}  
  
\def\tgb{\bar{\tau}}

\def\bC{\mbox{\boldmath $C$}}  
\def\bG{\mbox{\boldmath $G$}}  
\def\bH{\mbox{\boldmath $H$}}  
\def\bK{\mbox{\boldmath $K$}}  
\def\bM{\mbox{\boldmath $M$}}  
\def\bN{\mbox{\boldmath $N$}}  
\def\bO{\mbox{\boldmath $O$}}  
\def\bQ{\mbox{\boldmath $Q$}}  
\def\bR{\mbox{\boldmath $R$}}  
\def\bS{\mbox{\boldmath $S$}}  
\def\bT{\mbox{\boldmath $T$}}  
\def\bU{\mbox{\boldmath $U$}}  
\def\bV{\mbox{\boldmath $V$}}  
\def\bZ{\mbox{\boldmath $Z$}}  
  
\def\bcalS{\mbox{\boldmath $\mathcal{S}$}}  
\def\bcalG{\mbox{\boldmath $\mathcal{G}$}}  
\def\bcalE{\mbox{\boldmath $\mathcal{E}$}}  
  
\def\bgG{\mbox{\boldmath $\Gamma$}}  
\def\bgL{\mbox{\boldmath $\Lambda$}}  
\def\bgS{\mbox{\boldmath $\Sigma$}}  
  
\def\bgr{\mbox{\boldmath $\rho$}}  
  
\def\a{\alpha}  
\def\b{\beta}  
\def\g{\gamma}  
\def\G{\Gamma}  
\def\d{\delta}  
\def\D{\Delta}  
\def\e{\epsilon}  
\def\ve{\varepsilon}  
\def\z{\zeta}  
\def\h{\eta}  
\def\th{\theta}  
\def\k{\kappa}  
\def\l{\lambda}  
\def\L{\Lambda}  
\def\m{\mu}  
\def\n{\nu}  
\def\x{\xi}  
\def\X{\Xi}  
\def\p{\pi}  
\def\P{\Pi}  
\def\r{\rho}  
\def\s{\sigma}  
\def\S{\Sigma}  
\def\t{\tau}  
\def\f{\phi}  
\def\vf{\varphi}  
\def\F{\Phi}  
\def\c{\chi}  
\def\w{\omega}  
\def\W{\Omega}  
\def\Q{\Psi}  
\def\q{\psi}  
  
\def\ua{\uparrow}  
\def\da{\downarrow}  
\def\de{\partial}  
\def\inf{\infty}  
\def\ra{\rightarrow}  
\def\bra{\langle}  
\def\ket{\rangle}  
\def\grad{\mbox{\boldmath $\nabla$}}  
\def\Tr{{\rm Tr}}  
\def\Re{{\rm Re}}  
\def\Im{{\rm Im}}  

\title{Kadanoff-Baym approach to time-dependent quantum transport in AC and DC fields}

\author{Petri My\"oh\"anen$^1$, Adrian Stan$^1$, Gianluca Stefanucci$^{2,3}$ and Robert van Leeuwen$^{1,3}$ }

\address{$^1$Department of Physics, Nanoscience Center, FIN 40014, University of Jyv\"askyl\"a,
Jyv\"askyl\"a, Finland}
\address{$^2$Dipartimento di Fisica, Universit\`a di Roma Tor Vergata, Via della Ricerca Scientifica 1, I-00133 Rome, Italy}
\address{$^3$European Theoretical Spectroscopy Facility (ETSF)}

\ead{petri.myohanen@jyu.fi}

\begin{abstract}
We have developed a method based on the embedded Kadanoff-Baym 
equations to study the time evolution of open and inhomogeneous systems.
The equation of motion for the Green's function on the Keldysh contour 
is solved using different conserving many-body approximations for the 
self-energy. Our formulation incorporates
basic conservation laws, such as particle conservation, and
includes both initial correlations and initial embedding effects, 
without restrictions on the time-dependence of the external 
driving field.
We present results for the time-dependent density, current and dipole moment 
for a correlated tight binding chain connected to one-dimensional 
non-interacting leads exposed to DC and AC biases of various forms. 
We find that the self-consistent 
2B and GW approximations are in extremely good agreement with each other at all times,
for the long-range interactions that we consider.
In the DC case we show that the oscillations in the transients can be 
understood from interchain and lead-chain transitions in the system
and find that the dominant frequency corresponds to the HOMO-LUMO 
transition of the central wire.
 For AC biases with 
odd inversion symmetry odd harmonics to high harmonic order 
in the driving frequency are observed in the dipole moment, whereas for asymmetric applied bias
also even harmonics have considerable intensity.
In both cases we find that the HOMO-LUMO transition strongly mixes with the harmonics
leading to harmonic peaks with enhanced intensity at the HOMO-LUMO transition energy. 
\end{abstract}

\section{Introduction}

The progress in miniaturization of electronic devices has led to a completely 
new field of research, known as molecular electronics, in which it has become possible to prepare 
samples and to measure steady state conductivity properties of systems 
with single molecules squeezed between conducting electrodes \cite{H2nature,science}.
However, in future applications of nanodevices one is not only interested in their
steady properties, but one also wants to manipulate these systems in time.
For the control and operation of such devices 
the understanding of the time-dependent properties, such as switch-on 
times, currents, density fluctuations etc., becomes crucially important. In contrast 
to the steady-state approach, understanding these phenomena demands truly 
real-time treatment of the problem which is the topic of this paper.
\\
The study of periodic pulses is of special importance to manipulate 
and control the electron flow in nanoscale devices. In this paper we 
present results obtained from the solution of the Kadanoff-Baym (KB) 
equations for periodic voltage pulses. 
Within this framework also
 periodic gate 
voltages or barriers can be included without any additional computational effort. 
All these kind of signals can be experimentally 
realized and have been used to pump the electron current in several 
nanoscale structures ranging from molecules and quantum dots
\cite{kjvhf.1991,smcg.1999} to nanotubes \cite{lbtsajwc.2005,zgsgm.2008}.
The AC transport properties of nanoscale systems have been discussed 
recently \cite{tcd.2002,bsin.2004,mgm.2007,mmg.2009} and moreover, the advantages of real-time approaches in the 
context of quantum transport through molecular junctions driven out of equilibrium by 
periodic fields have been discussed in detail in Ref. \cite{skrg.2008}.

There is another reason for studying AC fields in quantum transport.
Usually the assumption is made that the external field in the leads is instantaneously screened.
However, in time-dependent transport transient times 
can be of the same order as the plasma oscillation period (see e.g. Ref.\cite{bsin.2004}). 
In such a case it would be reasonable to assume that at Hartree mean-field level the  
bias in the leads can be described by an AC field that alternates with the plasma frequency.
\\

In this article we use a recently proposed theoretical framework \cite{mssvl.2008, mssvl.2009} 
to study time-dependent electron transport through a quantum wire connected to two one-dimensional metallic 
leads in the presence AC and DC fields.  
The theoretical approach is based on the real-time propagation
of the KB  equations \cite{dsvl.2006,dvl.2007,vfva.2009,bbvlds.2009,book2,d.1984,kb.2000}
for open and interacting systems. The electron-electron interactions are 
handled perturbatively through the many-body self-energy for which we have implemented 
the HF (Hartree Fock), 2B (Second Born) and GW conserving 
approximations while the leads are taken into account through embedding self-energies.
We can calculate a wealth of
 observables such as currents, densities and dipole moments,
 and  study  electronic and 
initial correlations induced by the time nonlocal self-energy.

\section{Theory}
\label{theory}
\begin{figure}[t]
\centering
\includegraphics[width=1.0\textwidth]{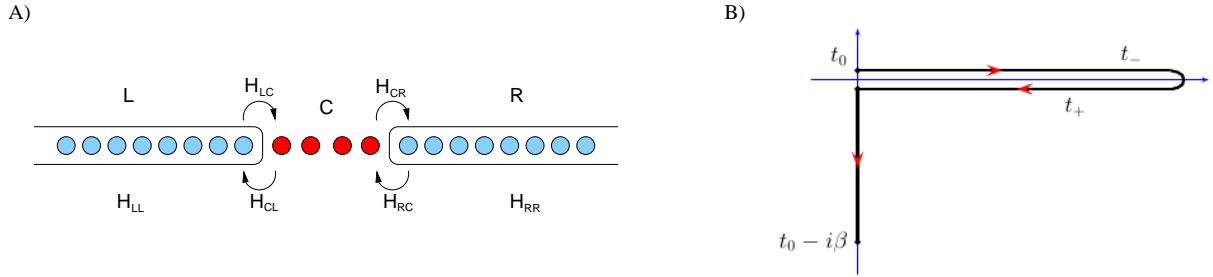}
\caption[]{A) The transport setup under consideration: A correlated central region is coupled to 
semi-infinite left and right noninteracting tight-binding leads. 
B) The Keldysh contour $\gamma$. Times on the upper/lower real-time 
branches 
are denoted with $t_{\mp}$. The imaginary (thermal) track runs from $t_0$ to 
$t_0 - i\beta$, with $\beta$ the inverse temperature.}
\label{fig1}
\end{figure}
\subsection{The Hamiltonian}
We consider a quantum correlated open system (a central region) 
coupled to noninteracting electron reservoirs (leads) and described by 
the Hamiltonian (see Fig.\ref{fig1} A)
\be
\hH (t) = \hH_{\rm C} (t) + \sum_{\alpha=L,R} \hH_\alpha (t) + \hH_{T} - 
\mu \hN.
\label{eq:ham}
\ee
Here $\hH_{\rm C}$, $\hH_{\a}$ and $\hH_{T}$ are the central region, the lead $\alpha=L,R$ and the 
tunneling Hamiltonians respectively, and $\hN$ is the particle number operator coupled to the chemical potential $\mu$.
The explicit expressions for these Hamiltonians read
\beq
\hH_{\rm C} (t) &=&
\sum_{ij,\sigma} h_{ij} (t) \hat{d}_{i\sigma}^{\dagger}\hat{d}_{j\sigma} +
\frac{1}{2}\sum_{\substack{ijkl\\ \sigma\sigma'}} 
v_{ijkl}\hat{d}_{i\sigma}^{\dagger}\hat{d}_{j\sigma'}^{\dagger}\hat{d}_{k\sigma'}\hat{d}_{l\sigma}, 
\label{eq:twobody} \\
\hH_\alpha (t) &=& U_{\a}(t) \hN_\alpha + 
\sum_{ij,\sigma} h_{ij}^{\alpha}  \, 
\hat{c}_{i\sigma\alpha}^{\dagger}\hat{c}_{j\sigma \alpha} ,
\label{eq:hamlead} \\
\hH_T  &=& \sum_{ij,\sigma \alpha} V_{i,j\alpha} [ \hat{d}_{i\sigma}^{\dagger}\hat{c}_{j\sigma \alpha} + 
\hat{c}_{j\sigma \alpha}^{\dagger}\hat{d}_{i\sigma} ],
\eeq
where $\sigma,\sigma'$ are the spin indices, $\hat{d}^{\dagger},\hat{d}$ 
and $\hat{c}^{\dagger}$,$\hat{c}$ are the creation and annihilation 
operators in the device and lead region and $i,j$ 
label a complete set of one-particle states in the corresponding 
subspaces of the system. 
The one-body part  $h_{ij}(t)$  of the central region Hamiltonian
is in general time-dependent  to account for driving fields such as gate voltages
or pumping fields.  The terms $v_{ijkl}$ in the two-body part are the two-electron integrals of the Coulomb interaction. Furthermore, 
the one-body part $h_{ij}^\alpha$ of the lead Hamiltonian  describes 
the metallic leads and the tunneling Hamiltonian 
$\hH_T$ contains the lead couplings to the central region.
The system is driven out of equilibrium by a homogeneous time-dependent bias voltage $U_\alpha (t)$ coupled to 
the number operator
$\hN_\alpha = \sum_{i,\sigma} \hat{c}_{i\sigma\alpha}^{\dagger}\hat{c}_{i\sigma \alpha}$ for lead $\alpha$.
This  field
can be considered as the sum of the externally applied field and the resulting screening field of the metallic leads which 
could, for instance, represent an AC field describing a plasmon oscillation.

\subsection{Kadanoff-Baym equations for the Keldysh Green's function}
\label{keldeom}
We study a system that is in equilibrium at an inverse temperature $\beta$ 
at times $t\leq t_0$ and described by a Hamiltonian $\hH_0$. For $t > t_0$ the system is
driven out of equilibrium by an external bias voltage and evolves in time according
to the Hamiltonian $\hH(t)$. The Keldysh Green's function for this system is defined as the expectation value of the 
contour-ordered product of the creation and annihilation 
operators \cite{dvl.2007,d.1984,w.1991,sa.2004} 
\be
\bcalG_{rs}(z,z') = -i \frac{\Tr \left\{ 
\mathcal{T} [e^{-i\int d\bar{z}\hH(\bar{z})}\hat{a}_r(z)\hat{a}^{\dagger}_s(z')] \right\}}{\Tr \left\{ e^{-\beta \hH_0}\right\}},
\label{kgreen}
\ee
where $\hat{a}$ and $\hat{a}^\dagger$ are either lead or central region operators and the indices $r$ and $s$ are
collective indices for position and spin. 
Furthermore, $z$ is a contour time variable and $\mathcal{T}$ orders the 
operators along the Keldysh contour (see Fig.\ref{fig1} B) by arranging the operators with later contour times to the left. 
The trace is taken over the many-body states of the system.
All the time-dependent one-particle properties can be calculated from the Green's 
function, for example the time-dependent density matrix is given by
\be
n_{rs} (t) = -i \bcalG_{rs} (t_{-},t_{+}),
\label{tddm}
\ee
where the time arguments $t_{\pm}$ are located on the lower/upper branch of the Keldysh contour.
Starting from Eq. (\ref{kgreen}) the equation of motion for the Green's function and its 
adjoint can easily be derived and are given by
\begin{eqnarray}
 i\de_z  \bcalG(z,z') &=& \delta(z,z')\mathbf{1} + \mathbf{H}(z) \bcalG(z,z')  
+ \int_{\gamma} d\bar{z}\,\bgS^{\rm MB}(z,\bar{z})\bcalG(\bar{z},z'),
\label{eqofmotion} \\
-i\de_{z'}  \bcalG(z,z') &=& \delta(z,z')\mathbf{1} + \bcalG(z,z')   \mathbf{H}(z')
+ \int_{\gamma} d\bar{z}\,\bcalG(z,\bar{z}) \bgS^{\rm MB}(\bar{z},z'),
\end{eqnarray}
where $\bgS^{\rm MB}$ is the many-body self-energy, $\mathbf{H}(z)$ is the one-body part of the full Hamiltonian 
and the integration is performed over the Keldysh-contour. 
These equations of motion need to be solved with the Kubo-Martin-Schwinger (KMS) boundary 
conditions: $\mbox{\boldmath $\mathcal{G}$}(t_0,z')=-\mbox{\boldmath $\mathcal{G}$}(t_0-i\beta,z')$ 
and $\mbox{\boldmath $\mathcal{G}$}(z,t_0)=-\mbox{\boldmath $\mathcal{G}$}(z,t_0-i\beta)$.
All the quantities are considered as block matrices with a structure given by
\begin{equation}\label{1bham}
\mathbf{H} = 
\left[
\begin{array}{ccc}
\mathbf{H}_{\textnormal{\scriptsize{LL}}} & 
\mathbf{H}_{\textnormal{\scriptsize{LC}}} &  \mathbf{0} \\
\mathbf{H}_{\textnormal{\scriptsize{CL}}} & 
\mathbf{H}_{\textnormal{\scriptsize{CC}}} & 
\mathbf{H}_{\textnormal{\scriptsize{CR}}} \\
\mathbf{0} & \mathbf{H}_{\textnormal{\scriptsize{RC}}} & 
\mathbf{H}_{\textnormal{\scriptsize{RR}}}
\end{array}
\right]
\,\,,\,\,
\bgS^{\rm{MB}} = 
\left[
\begin{array}{ccc}
\mathbf{0} & \mathbf{0} & \mathbf{0}\\
\mathbf{0} & \bgS^{\rm{MB}}_{\rm{CC}} [\bcalG_{\rm{CC}}] & \mathbf{0} \\
\mathbf{0} & \mathbf{0} & \mathbf{0}
\end{array}
\right]
\,\,,\,\,
\mbox{\boldmath $\mathcal{G}$} = 
\left[
\begin{array}{ccc}
\bcalG_{\textnormal{\scriptsize{LL}}} &\bcalG_{\textnormal{\scriptsize{LC}}} & \bcalG_{\textnormal{\scriptsize{LR}}}\\
\bcalG_{\textnormal{\scriptsize{CL}}} & \bcalG_{\textnormal{\scriptsize{CC}}} & \bcalG_{\textnormal{\scriptsize{CR}}} \\
\bcalG_{\textnormal{\scriptsize{RL}}} &\bcalG_{\textnormal{\scriptsize{RC}}} & \bcalG_{\textnormal{\scriptsize{RR}}}
\end{array}
\right],
\end{equation}
where the different block matrices describe the projections onto different subregions. 
We assume that there is no direct coupling between the leads. 
The many-body self-energy has nonzero elements only in the central region. This follows immediately from
the diagrammatic expansion of the self-energy.
More specifically, if we expand the self-energy in powers of the two-body interactions we see that the interaction matrix
elements $v_{ijkl}$ only connect sites in the central region and therefore all Green function lines in the corresponding Feynman diagrams
are of the type $\bcalG_{\rm{CC}}$.
Therefore $\bgS^{\rm{MB}} [\bcalG_{\rm{CC}}]$ is a functional of $\bcalG_{\rm{CC}}$ only. Note that we do not expand
in the lead-device couplings which are all exactly incorporated in the one-body matrix $\mathbf{H}$.
To study the dynamical processes and to extract the observables of interest such as the densities and currents 
in the whole open and connected system we need the solution for the full Green function in different subregions of the system.
In particular we need the Green's functions $\bcalG_{\rm{CC}}$ projected onto central region C and $\bcalG_{\alpha\alpha}$ projected
onto the lead $\alpha$ region. For these we need to extract from the block matrix 
structure of the Green's function an equation for $\bcalG_{\rm{CC}}$ and $\bcalG_{\alpha\alpha}$.
The projection of the equation of motion (\ref{eqofmotion}) onto 
regions ${\rm CC}$ and $\alpha {\rm C}$ gives
\begin{equation}\label{eomCC}
\begin{split}
&\Bigl\{ i\de_z\mathbf{1} - \mathbf{H}_{\rm CC}(z) \Bigr\}\bcalG_{\rm CC}(z,z') = 
\delta(z,z')\mathbf{1}\,\, +
\sum_{\alpha}\mathbf{H}_{{\rm C}\alpha}\bcalG_{\alpha\rm C}(z,z') 
+ \int_{\gamma} d\bar{z}\,\bgS_{\rm CC}^{\rm MB}(z,\bar{z})\bcalG_{\rm CC}(\bar{z},z'),
\end{split}
\end{equation}
for the central region and
\be
\Bigl\{ i\de_z\mathbf{1} - \mathbf{H}_{\alpha \alpha}(z) \Bigr\} \bcalG_{\alpha \rm{C}} (z,z') = 
 \mathbf{H}_{\alpha \rm{C}}\bcalG_{\rm{CC}}(z,z') ,
\label{eomCa}
\ee
for the projection on the interface $\alpha {\rm C}$ region. We see that the right hand side of Eq.(\ref{eomCa})
is of a simple form as a result of the absence of two-body interactions in the leads.
The uncontacted and biased lead Green's function $\bg_{\alpha \alpha}$ satisfies the equation
of motion
\be
\Bigl\{i\partial_z\mathbf{1} - \mathbf{H}_{\alpha \alpha}(z)\Bigr\} \bg_{\alpha \alpha}(z,z') 
=\delta(z,z')\mathbf{1}.
\label{smallg}
\ee
Using this equation we can construct the solution for
$\bcalG_{\alpha \rm{C}}$ and it reads
\begin{equation}
\bcalG_{\alpha\rm C}(z,z') =  
\int_{\gamma} d\bar{z}\,\bg_{\alpha\alpha}(z,\bar{z})\,
\mathbf{H}_{\alpha\rm C}\bcalG_{\rm CC}(\bar{z},z').
\label{GalphaC}
\end{equation}
Taking into account Eq. (\ref{GalphaC}) the second term on the righthand side of Eq. (\ref{eomCC}) becomes
\be
\sum_{\alpha} \mathbf{H}_{\rm{C}\alpha} \bcalG_{\alpha \rm{C}} (z,z') =
\int_{\gamma} d\bar{z} \,\bgS_{\rm{em}} (z,\bar{z}) \bcalG_{\rm{CC}} 
(\bar{z},z'),
\label{sig_em}
\ee
where we have introduced the \emph{embedding} self-energy
\be
\bgS_{\rm{em}} (z,z') = \sum_\alpha \bgS_{\rm{em},\alpha} (z,z') 
 = \sum_{\alpha}\mathbf{H}_{{\rm C}\alpha}\,
\bg_{\alpha\alpha}(z,z') 
\mathbf{H}_{\alpha\rm C},
\label{embedding}
\ee
accounting for the tunneling of electrons between the central region and the leads. 
As we see from the definition the embedding self-energy 
$\bgS_{\rm{em},\alpha}$ depends
only on the coupling Hamiltonians and on the isolated lead Green's 
function $\bg_{\alpha\alpha}$. Furthermore, the Green's function
is determined once the isolated lead Hamiltonian $\hH_\alpha$ of Eq. (\ref{eq:hamlead}) is specified.
Inserting (\ref{sig_em}) back to (\ref{eomCC}) then gives the equation of motion
\begin{equation}
\begin{split}
&\Bigl\{ i\de_z\mathbf{1} - \mathbf{H}_{\rm CC}(z) \Bigr\} 
\bcalG_{\rm CC}(z,z')=\delta(z,z')\mathbf{1} + \int_{\gamma} d\bar{z}\,\left[ 
\bgS_{\rm CC}^{\rm MB}+
\bgS_{\rm em}\right](z,\bar{z})\,\bcalG_{\rm CC}(\bar{z},z').
\end{split}
\label{embeddedEOM}
\end{equation}
An adjoint equation can be derived similarly \cite{mssvl.2009}.
Equation (\ref{embeddedEOM}) is an exact equation 
for the Green's function $\bcalG_{\rm{CC}}$ for the class of Hamiltonians of Eq. (\ref{eq:ham}),
provided that an exact expression for $\bgS^{\rm{MB}}_{\rm CC}[\bcalG_{\rm{CC}}]$ 
as a functional of $\bcalG_{\rm{CC}}$ is inserted.

To apply Eq. (\ref{embeddedEOM}) in practice we need to transform it to real-time equations
that we solve by time-propagation.
This can be done in Eq. (\ref{embeddedEOM})
by considering time-arguments of the Green's function and self-energy on different branches of the contour.
We therefore have to define these components first.
Let us thus consider a function on the Keldysh contour of the general form
\be
F(z,z') = F^\delta (z)\delta (z,z') 
 + \theta (z,z') F^> (z,z') + \theta (z',z) F^< (z,z') ,
\ee
where $\theta (z,z')$
is a contour Heaviside function \cite{d.1984}, i.e., $\theta (z,z')=1$
for $z$ later than $z'$ on the contour and zero otherwise,
and $\delta (z,z')=\partial_z \theta (z,z')$ is the contour delta function.
By restricting the variables $z$ and $z'$ on different branches of the contour
we can define the various
components of $F$ as
\beq
F^\lessgtr (t,t') &=& F(t_{\mp},t_{\pm}'), \\
F^{\rceil} (t,\tau) &=& F(t_{\pm},t_0-i\tau), \\
F^{\lceil} (\tau,t) &=& F(t_0-i\tau,t_{\pm}) ,\\
F^M (\tau - \tau') &=& -i F(t_0-i\tau, t_0-i\tau'),
\eeq
and 
\be
F^{R/A} (t,t') = F^\delta (t) \delta (t-t') \pm \theta (\pm t \mp t') [F^>(t,t')-F^< (t,t')].
\ee
For the Green's function there is no singular contribution, i.e., $\bcalG^\delta=0$, but the
self-energy has a singular contribution of Hartree-Fock form, i.e.,
$\bgS^\delta = \bgS^{\rm HF} [\bcalG]$ \cite{d.1984}. 
With these definitions we can now convert Eq. (\ref{embeddedEOM}) into
equations for the separate components. This is conveniently done using the conversion table
in Ref. \cite{TDDFTbook}. This leads to
a set of coupled real-time differential equations, known as the Kadanoff-Baym equations 
\beq
i\de_t \bcalG^\lessgtr(t ,t') &=& \mathbf{H}_{\rm{CC}}(t) \bcalG^\lessgtr(t,t')
+\left[\bgS^{R} \cdot \bcalG^\lessgtr 
+ \bgS^\lessgtr \cdot \bcalG^A  + \bgS^\rceil \star \bcalG^\lceil \right](t,t'), 
\label{eq:kb1} \\
-i\de_{t'} \bcalG^\lessgtr(t, t') &=&  \bcalG^\lessgtr(t,t') \mathbf{H}_{\rm{CC}}(t')
+\left[\bcalG^{R} \cdot \bgS^\lessgtr + \bcalG^\lessgtr \cdot \bgS^A + \bcalG^\rceil \star \bgS^\lceil \right](t,t'), 
\label{eq:kb2} \\
i\de_t\bcalG^\rceil(t,\tau) &=& \mathbf{H}_{\rm CC}(t) \bcalG^\rceil(t,\tau)
+\left[\bgS^R \cdot \bcalG^\rceil + \bgS^\rceil \star \bcalG^M \right](t,\tau),
\label{eq:kb3} \\
-i\de_t\bcalG^\lceil(\tau,t) &=&  \bcalG^\lceil(\tau,t) \mathbf{H}_{\rm CC}(t)
+\left[\bcalG^\lceil \cdot \bgS^A + \bcalG^M \star \bgS^\lceil \right](\tau,t),
\label{eq:kb4} \\
- \de_{\tau} \bcalG^M (\tau-\tau') &=& \mathbf{1} \delta (\tau-\tau') + \mathbf{H}_{\rm{CC}} \bcalG^M (\tau-\tau')
+i \left[\bgS^M \star \bcalG^M \right](\tau-\tau').
\label{eq:Matsubara}
\eeq 
Here the self-energy is the sum of the many-body and embedding self-energies. The superscript notation 
for the various components of self-energy and the Green's function, "$M$",$<$,$>$,$\rceil$ and $\lceil$ 
are used to identify the time-arguments on different parts of the 
Keldysh contour \cite{mssvl.2009,sa.2004,TDDFTbook}. The notations 
$\cdot$ and $\star$ denote the real-time and imaginary-time convolutions:
\begin{equation}
\left[ a\cdot b \right](t,t') = \int_0^{\infty} a(t,\bar{t})b(\bar{t},t') d\bar{t} \quad , \quad
\left[ a\star b \right](t,t') = -i \int_0^{\beta} a(t,\tau)b(\tau,t') d\tau.
\end{equation}
In practice, the Matsubara equation (\ref{eq:Matsubara}) is solved first since it is disconnected from the 
real-time KB equations. The real-time Green's functions are then initialized with the Matsubara Green's function 
as: $\bcalG^> (0,0) = i \bcalG^M (0^+)$, $\bcalG^< (0,0) = i \bcalG^M (0^-)$, $\bcalG^\rceil (0,\tau) = i \bcalG^M (-\tau)$ 
and $\bcalG^\lceil (\tau,0) = i \bcalG^M (\tau)$ and then the equations (\ref{eq:kb1})-(\ref{eq:kb4}) are solved
using the time-propagation method described in Ref. \cite{sdvl.2009b}.

\subsection{Conserving approximations}
\begin{figure}[t]
\centering
\includegraphics[width=0.6\textwidth]{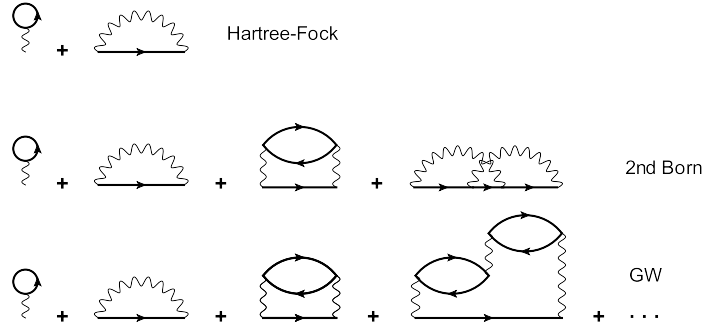}
\caption{Diagrammatic representation of the self-energies for the conserving many-body approximations used. }
\label{figDiagrams}
\end{figure}

The importance of particle number conservation in quantum transport transport has been carefully addressed in Refs. 
\cite{tr.2008,t.2008}.
The particle number conservation law in open systems is given by
\begin{equation}
\frac{d N_{\rm C}(t)}{d t} =I_{\rm L}(t) +I_{\rm R} (t)\label{cc3},
\end{equation}
and expresses the fact that the change in the number of particles in the central region sums up to
the total current that flows into the leads.
Particle number conservation is guaranteed within the KB
approach by using expressions for the self-energy that can be obtained
as functional derivative with respect to the Green's function of a functional $\Phi$: 
\be
\bgS^{\rm{MB}}_{{\rm CC},rs} [\bcalG_{\rm{CC}}](z,z') = 
\frac{\delta \Phi [\bcalG_{\rm{CC}}]}{\delta \bcalG_{{\rm CC},sr} (z',z)}.
\label{eq:Phifunc}
\ee
Baym has proven~\cite{b.1962} that such $\Phi$-derivable or conserving approximations \cite{bk.1961,b.1962,vbdvls.2005} automatically lead
to satisfaction of the conservation laws provided that the equations of motion for the Green's function are solved 
to full self-consistency.
Commonly used conserving approximations for $\bgS^{\rm{MB}}_{\rm CC}[\bcalG_{\rm{CC}}]$ are the
Hartree-Fock, second Born and GW approximations \cite{dvl.2007,sdvl.2009b,sdvl.2009,dvl.2005}
which are displayed diagrammatically in Fig.{\ref{figDiagrams}}.

\subsection{Time-dependent current}
\label{chcons}

An equation for the time-dependent current flowing into the lead $\alpha$ can be derived from the
time-derivative of the number of particles in lead $\alpha$ using the equation of motion for the
lead Green's function $\bcalG_{\alpha \alpha}$ \cite{mssvl.2009}.
This yields
\be
I_{\alpha}(t) = 2\textnormal{Re}\,\Tr_{\rm C}\left[ \bcalG_{\rm C \alpha}^{<}(t,t)H_{\alpha C} \right].
\ee
If we insert the adjoint of Eq. (\ref{GalphaC}) and extract the different components from the resulting equation
using the conversion table in Ref.\cite{TDDFTbook}, the equation for the current becomes \cite{mssvl.2009}
\beq
I_{\alpha}(t)
&=&2\textnormal{Re}\,\Tr_{\rm C} \left[ 
\int_{t_0}^{t}d\bar{t}\,
\bcalG_{\rm CC}^{<}(t,\bar{t}) \bgS_{{\rm em},\a}^{A}(\bar{t},t) 
+\int_{t_0}^{t}d\bar{t}\,
\bcalG_{\rm CC}^{R}(t,\bar{t}) \bgS_{{\rm em},\a}^{<}(\bar{t},t) \right.
 \nonumber \\
&& \left. -i\int_{0}^{\b}d\t\,  \bcalG_{\rm CC}^{\rceil}(t,\t)\bgS_{{\rm em},\a}^{\lceil}(\t,t) \right].
\label{current2}
\eeq
The first two terms in this equation contain integrations over earlier times from $t_0$ to $t$ and
take into account the nontrivial memory effects arising from the
time-nonlocality of the embedding self-energy and the Green's functions. Furthermore,
the last integral over the imaginary track incorporates the initial many-body and embedding effects. 
Equation (\ref{current2}) provides a generalization of the widely used 
Meir-Wingreen formula \cite{mw.1992,jwm.1994} to the transient time-domain \cite{mssvl.2009}.

\subsection{Nonequilibrium spectral function}

The nonequilibrium spectral function for real times is defined as
\be
A(t,t') = i\,\Tr_{\rm C}[\bcalG_{\rm CC}^> - \bcalG_{\rm CC}^<](t,t') =\sum_{i \in C,\sigma} \langle \left\{ \hat{d}_{i\sigma} (t) , \hat{d}_{i\sigma}^\dagger (t') \right\} \rangle .
\ee
where the brackets in the last term denote the anti-commutator.
It is common to Fourier transform this function with respect to the relative time-coordinate $\tau=t-t'$
for a given value of the average time $T=(t+t')/2$.
The spectral function in the new coordinates is then 
\be
A(T,\omega) = - \Im  \int \frac{d\t}{2 \pi} \, 
e^{i \omega  \t}  \Tr_{\rm C} [ \bcalG_{\rm CC}^> - \bcalG_{\rm CC}^<] (T+ 
\frac{\t}{2}, T -  \frac{\t}{2}) = \Re  \int \frac{d\t}{2 \pi} \, e^{i \omega  \t} A(T+\frac{\tau}{2},T-\frac{\tau}{2})
\ee
If the time-dependent external field becomes constant after some switching time, then also
the spectral function becomes
independent of $T$ after some transient period. In that case we will denote the spectral function by
$A(\omega)$.
The spectral function for a system in equilibrium has peaks at the addition and removal energies.
In the nonequilibrium case this is less clear as one can, in general, not use a Lehmann expansion
to prove this fact. However, for a system to which a bias is suddenly applied one can still show that
the spectral function will contain peaks at the addition and removal energies from the various excited
states of the biased system.

\subsection{The density in the leads}
The density in the leads can be obtained from the Green's function $\bcalG_{\alpha \alpha}$.
This Green's function satisfies an equation of motion that can be obtained by
projecting the equation (\ref{eqofmotion}) into the $\alpha\alpha$ region:
\begin{equation}\label{eomaa}
\begin{split}
&\Bigl\{ i\de_z\mathbf{1} - \mathbf{H}_{\alpha\alpha}(z) \Bigr\}\bcalG_{\alpha\alpha}(z,z') = 
\delta(z,z')\mathbf{1}\,\, + \mathbf{H}_{{\alpha \rm C}}\bcalG_{\rm C \alpha}(z,z'). 
\end{split}
\end{equation}
Using the adjoint of (\ref{GalphaC}) in (\ref{eomaa}) we have
\beq
\Bigl\{ i \de_z - \mathbf{H}_{\a \a} (z)\Bigl\} \bcalG_{\a \a} (z,z') &=& \delta (z,z') \mathbf{1} +\int_{\gamma} d \bar{z} \bgS_{\rm{in},\alpha} (z,\bar{z}) \bg_{\a \a} 
(\bar{z},z') ,
\label{eq:Gaa_eom}
\eeq
where we defined the {\em inbedding} self-energy as
\be
\bgS_{\rm{in},\alpha} (z,z') = \mathbf{H}_{\a {\rm C}} 
\bcalG_{\rm{CC}} (z,z') \mathbf{H}_{{\rm C} \a}.
\ee
Equation (\ref{eq:Gaa_eom}) can now be solved using Eq.(\ref{smallg}). 
By taking
the time arguments at $t_\pm$ we get
\beq
\bcalG_{\a \a} (t_-,t_+) &=& \bg_{\a \a} (t_-,t_+) +
\int_{\gamma} d \bar{z} d \bar{\bar{z}} \bg_{\a \a} (t_-,\bar{z}) \bgS_{\rm{in},\a} 
(\bar{z},\bar{\bar{z}}) \bg_{\a \a} (\bar{\bar{z}},t_+).
\label{leaddens}
\eeq
We see from Eq.(\ref{tddm}) that from this equation we can extract the time-dependent spin occupation of orbital $i$ in lead $\alpha$ 
by taking $r=s=i\sigma \alpha$.
The first term on the r.h.s. of Eq. (\ref{leaddens}) gives
the density of uncontacted biased lead $\alpha$ and the second term gives a correction
to the density induced by the presence of the correlated scattering region. 
In practice we first solve the Kadanoff-Baym equations for $\bcalG_{\rm{CC}}$
and then we construct the inbedding self-energy $\bgS_{\rm{in}}$
which is then used to calculate the time-dependent lead 
density according to (\ref{leaddens}).\\

\subsection{Embedding self-energy}
\label{emdse}
In this section we derive the expression for the embedding self-energy for the case of 
one-dimensional noninteracting tight-binding leads. From equation (\ref{embedding}) we see 
that the embedding self-energy is given by
\be
\bgS_{{\rm em},\alpha,kl} (z,z') = \sum_{ij}^{\mathcal{N_\a}} V_{k,i\alpha} \bg_{\a 
\a,ij} (z,z') V_{j\alpha,l},
\label{embedding2}
\ee
where $k$,$l$ label the sites in the central region and $i$,$j$ the sites in the lead $\a$.
Furthermore, $\mathcal{N}_\a$ is the number of sites in the lead $\a$
(at the end of the derivation we take $\mathcal{N}_\a \rightarrow \infty$).
We now express the biased lead Green's functions using the eigenbasis of the biased lead Hamiltonian: 
\beq
\bg_{\a\a,ij}^{\lessgtr}(z,z') &=& \sum_{pq}^{\mathcal{N}_\a} U^\alpha_{ip}\tilde{\bg}^{\lessgtr}_{\a\a,pq}(z,z')U_{q j}^{\alpha \dagger},
\eeq
where
\beq
\tilde{\bg}_{\a\a,pq}^<(z,z')&=&i\delta_{pq}f(\epsilon_{p\alpha})e^{-i\int_{z'}^z (\epsilon_{p\alpha}-\mu+U_{\alpha}(\bar{z}))d\bar{z}},
\label{eq:gf1}\\
\tilde{\bg}_{\a\a,pq}^> (z,z')&=&i\delta_{pq}(f(\epsilon_{p\alpha})-1)e^{-i\int_{z'}^z (\epsilon_{p\alpha}-\mu+U_{\alpha}(\bar{z}))d\bar{z}},
\label{eq:gf2}
\eeq
are the lead Green's functions in the eigenbasis of the leads with eigenvalues $\epsilon_{p\alpha}$. 
The columns of matrix $\mathbf{U}^\a$ represent these eigenstates in site basis. The function
$f(\epsilon)=1/(e^{\beta (\epsilon-\m)}+1)$ is the Fermi distribution function.
Inserting Eqs.(\ref{eq:gf1}) and (\ref{eq:gf2}) into the definition of the embedding self-energy (\ref{embedding2}) and defining the
function
\be
\Gamma_{kl,\alpha}(\epsilon) = 2\pi\sum_{ijp}^{\mathcal{N}_\a}V_{k,i\alpha}U^\alpha_{ip}\delta(\epsilon-\epsilon_{p\alpha})U_{pj}^{\alpha \dagger}V_{j\alpha,l}
\ee
we get an expression for the one-dimensional lead self-energy
\beq
\bgS_{{\rm em},\alpha,kl}^<(z,z') &=& ie^{-i\int_{z'}^z U_{\alpha}(\bar{z})d\bar{z}}\int \frac{d\epsilon}{2\pi}f(\epsilon)\Gamma_{kl,\alpha}(\epsilon)e^{-i(\epsilon-\mu)(z-z')},\\
\bgS_{{\rm em},\alpha,kl}^>(z,z') &=& ie^{-i\int_{z'}^z U_{\alpha}(\bar{z})d\bar{z}}\int \frac{d\epsilon}{2\pi}(f(\epsilon)-1)\Gamma_{kl,\alpha}(\epsilon)e^{-i(\epsilon-\mu)(z-z')}.
\eeq
The remaining task is to evaluate the function $\Gamma_{kl,\alpha}(\epsilon)$. For a tight-binding (TB) chain we have
\beq
\epsilon_{p\alpha} &=& a^{\alpha} + 2T_{\alpha}\cos(\phi_p),\\
U_{np}^{\alpha} &=&\sqrt{\frac{2}{\mathcal{N}_\a+1}}\sin(n\phi_p),
\label{eigvals}
\eeq
where $a^{\alpha}$ and $T_{\alpha}$ are the tight binding on-site and hopping parameters for the lead $\alpha$ and 
$\phi_p = \frac{p\pi}{\mathcal{N}_\a+1}$, $p=1...\mathcal{N}_\a\,,n=1...\mathcal{N}_\a$.  We consider the case that the coupling Hamiltonian has nonzero elements only 
at the terminal points adjacent to the endpoints of the central region nanowire, that is when $i=j=1$.
The function $\Gamma_{kl,\alpha}(\epsilon)$ then becomes
\beq
\Gamma_{kl,\alpha}(\epsilon) = 2\pi V_{k,1\alpha}V_{1\alpha,l}\sum_{p=1}^{\mathcal{N}_\a} U^\alpha_{1p}\delta(\epsilon-\epsilon_{p\alpha})U_{p1}^{\alpha \dagger}.
\label{gamma}
\eeq
Inserting Eqs. (\ref{eigvals}) into (\ref{gamma}) and taking the $\mathcal{N}_\a \rightarrow \infty$ limit we obtain
\beq
\Gamma_{kl,\alpha}(\epsilon) = \frac{2 
V_{k,1\alpha}V_{1\alpha,l}}{|T_{\alpha}|}\sqrt{1-\left( \frac{\epsilon-a^{\alpha}}{2T_{\alpha}}\right)^2}
\times \theta(2|T_{\alpha}|-|\epsilon-a^{\alpha}|).
\label{gamma2}
\eeq
From the definition of $\Gamma_{kl,\alpha}(\epsilon)$ we see that it describes the weighted 
density of states of the lead $\alpha$ as a function of the energy. 
Equation (\ref{gamma2}) then yields the expression for the 1D lead self-energy
\be
\bgS_{{\rm em},\alpha,kl}^<(z,z') = 
i\frac{V_{k,1\alpha}V_{1\alpha,l}}{\pi |T_{\alpha}|}
e^{-i\int_{z'}^z 
U_{\alpha}(\bar{z})d\bar{z}}\int_{a^{\alpha}-2|T_{\alpha}|}^{a^{\alpha}+2|T_{\alpha}|} 
d\epsilon f(\epsilon)\sqrt{1-\left( \frac{\epsilon-a^{\alpha}}{2T_{\alpha}}\right)^2} e^{-i(\epsilon-\mu)(z-z')},
\ee
and the $\bgS_{{\rm em},\alpha,kl}^>$ component is obtained by simply replacing the $f(\epsilon)$ with $f(\epsilon)-1$. Furthermore, 
the $\lceil$,$\rceil$ and "M" components are similarly obtained by considering the time-arguments on different parts of the Keldysh contour.

\section{Numerical results}
\label{numres}

We now apply the KB approach to study the transport dynamics of a 
quantum wire consisting of 4 sites 
connected to left and right one-dimensional TB leads, see Fig. \ref{fig1} A.
The nearest neighbour hoppings in the leads are set to 
$T_{\rm L} = T_{\rm R} = -2.0$ and the on-site energy $a^{\a}$ is set equal to the chemical 
potential, $a^\alpha=\mu$, yielding a half-filling for the lead energy bands.
The values for the coupling Hamiltonian are set to $V_{1,1L} = V_{4,1R} = -0.5$ and  
the central region sites have the on-site energies $h_{ii}=0$ and hopping parameters
$h_{ij}=-1.0$ between nearest neighbour sites $i$ and $j$.
The electron-electron interaction in the central region has the 
form  $v_{ijkl} = 
v_{ij}\,\delta_{il}\delta_{jk}$ with 
\begin{equation}
v_{ij} =
\begin{cases}
v_{ii} & i=j\\
\frac{v_{ii}}{2|i-j|} & i\neq j 
\end{cases}
\label{eq:Coulomb}
\end{equation}
where $v_{ii} = 1.5$.
\begin{figure}[t]
\centering
\includegraphics[width=1.0\textwidth]{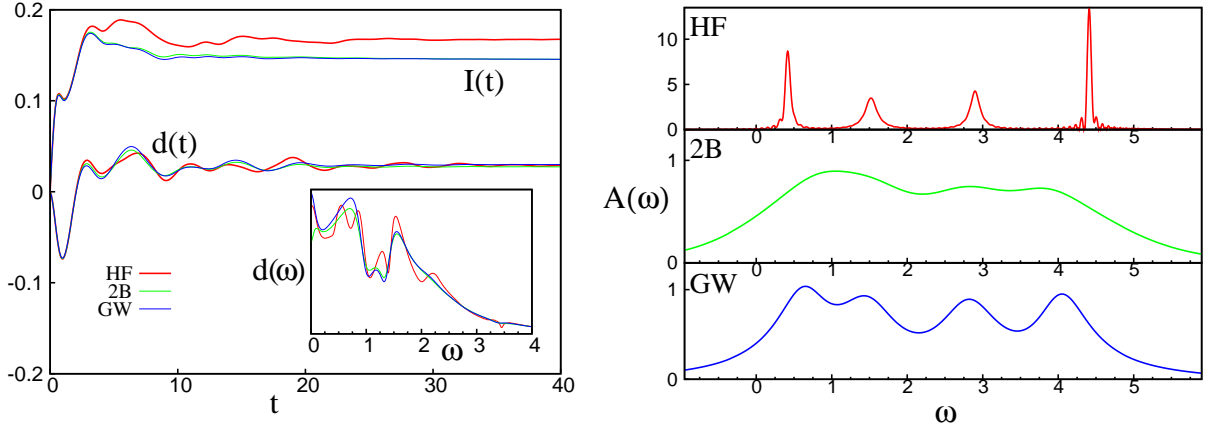}
\caption[]{Left panel: Time-dependent right current $I(t)$ and dipole moment 
$d(t)$ for different self-energy approximations when a DC bias 
$U_{0}=1.4$ is applied to the leads. The inset shows the 
absolute value Fourier transform of the dipole moment. Right panel: Steady state spectral 
function $A(\omega)$.}
\label{fig3}
\end{figure}
\begin{figure}[t]
\centering
\includegraphics[width=1.0\textwidth]{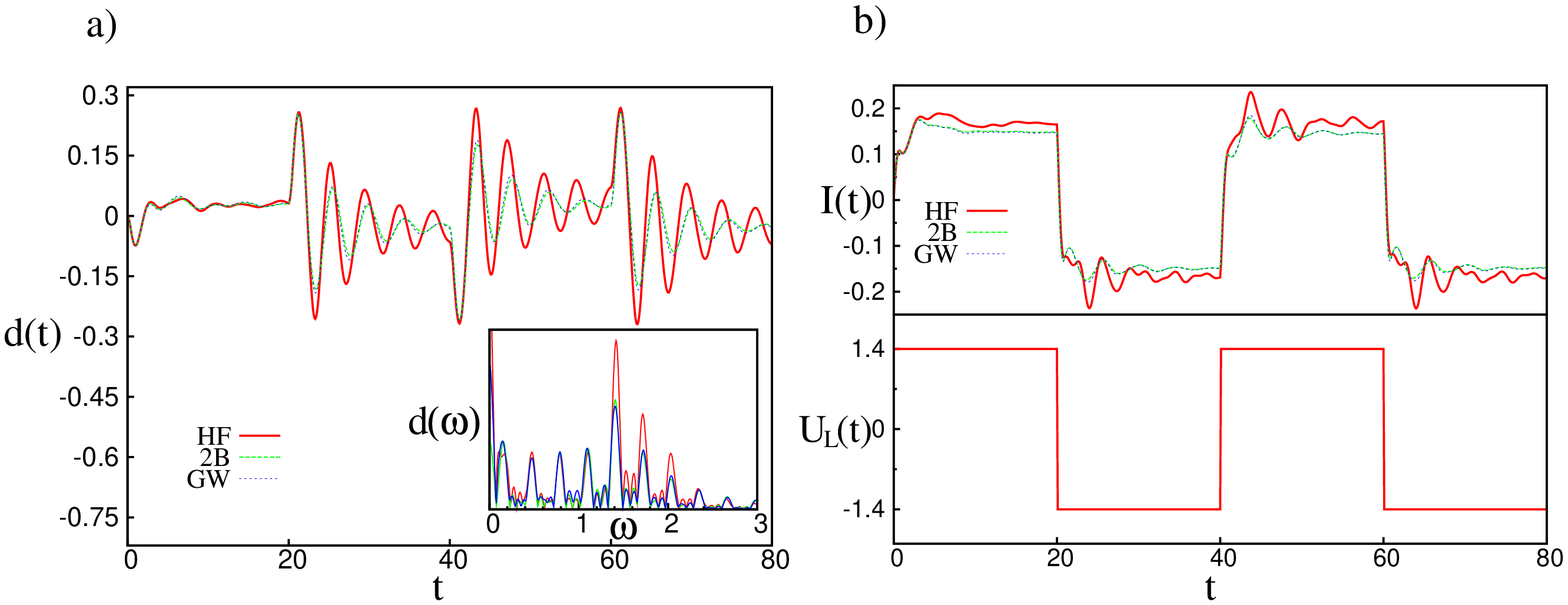}
\caption[]{Panel a): Dipole moment $d(t)$ corresponding to bias $U_1(t)$. The inset shows the absolute value of the Fourier 
transform of the dipole moment. Panel b): Time-dependent current flowing into the right lead
(top panel) and the time-dependent bias $U_1(t)$ (bottom panel).
}
\label{fig4}
\end{figure}
\begin{figure}[]
\centering
\includegraphics[width=1.0\textwidth]{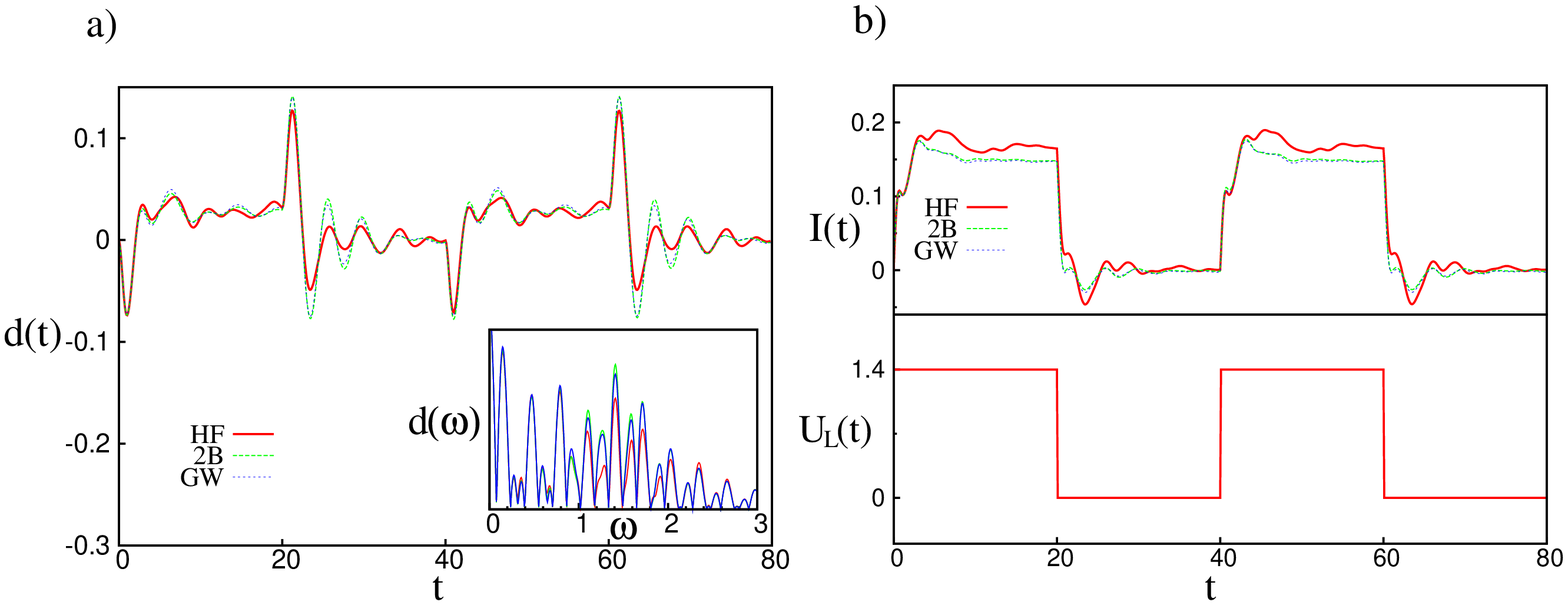}
\caption[]{Panel a): Dipole moment $d(t)$ corresponding to bias $U_2(t)$. The inset shows the absolute value of the Fourier 
transform of the dipole moment. Panel b): Time-dependent current flowing into the right lead
(top panel) and the time-dependent bias $U_2(t)$ (bottom panel).}
\label{fig5}
\end{figure}
The chemical potential is fixed between the highest occupied molecular orbital (HOMO) and lowest unoccupied molecular orbital (LUMO)
levels of the isolated central region and has the value $\mu = 2.26$.
Furthermore, the inverse temperature is set to $\beta=90$ corresponding to the zero temperature limit. 
We use different time-dependent bias voltages. We consider a sudden switch on DC case  of the form $U_{L}(t) = -U_{R}(t) = U_0 \theta(t-t_0)$ 
and periodic AC pulses $U_{L}(t) = -U_{R}(t)=U_{1,2}(t)$ of the forms 
\beq
U_1 (t) &=& 
\begin{cases}
U_0 & 0 \leq t < \frac{T_0}{2}\\
-U_0 & \frac{T_0}{2} \leq t < T_0\\
\end{cases}\,\,\,\,\,\, \textnormal{(block bias)}
\label{eq:block}\\
U_2 (t) &=& 
\begin{cases}
U_0 & 0 \leq t < \frac{T_0}{2}\\
0 & \frac{T_0}{2} \leq t < T_0\\
\end{cases}\,\,\,\,\,\,\,\,\,\,\, \textnormal{(on-off bias)}
\label{eq:onoff}
\eeq
which for $t > T_0$ are periodically extended using $U(t+T_0) = U(t)$.
The time $T_0$ is the period of the complete bias cycle corresponding to 
oscillation frequency $\omega_0 = 2\pi/T_0$.
In all cases we choose the amplitude of the bias voltage to be $U_0 = 1.4$ 
and for AC biases we set the period to be $T_0= 40$ a.u. 
Finally, to study how the charge distributes along the chain after a bias voltage is switched on
we calculate the time-dependent dipole moment
\be
d(t) = \sum_{i=1}^4 x_i \, n_i (t)
\ee
where the $x_i$ are the coordinates of the sites of the chain (with a lattice spacing of one),
with origin halfway between sites 2 and 3, and $n_i(t)$ is the site occupation number.\\
We start by considering the case of a suddenly switched DC bias.
In the left panel of Fig. \ref{fig3} we compare the time-dependent dipole moments (three lowermost curves) and
transient currents (three uppermost curves) for the HF, 2B and GW approximations. The right panel shows the steady-state spectral functions.
We see that the Hartree-Fock peaks are relatively sharp with a width approximately given by the
imaginary part of the embedding self-energy at the Fermi level of $\Gamma =2 V^2/T_\alpha\approx 0.25$ where $V=V_{1,1L} = V_{4,1R}$. On the other
hand the 2B and GW spectral functions are very broadened due to strongly enhanced quasi-particle scattering at finite bias \cite{t.2008,mssvl.2009}.
We see that the Hartree-Fock 
approximation produces a slightly larger current compared to the 2B and GW approximations.
This is in keeping with the fact that the HF spectral function integrated
over the bias window leads to a slightly larger value as compared to those from the 2B and GW approximations.
We further see that the correlated 2B and GW 
approximations lead to currents and dipole moments in close agreement 
with each other. 
This indicates that the dynamical screening 
of the Coulomb interaction as described by the first bubble diagram of the self-energy
has a leading contribution.\\
\\
In the inset of the left panel in Fig. \ref{fig3} we display the absolute value of the Fourier transform of the
dipole moment. We see a number of peaks that can be directly related to transitions from the lead electrochemical
potentials $\mu+U_\alpha$ to the lead levels and between levels in the central sites themselves.
The nature of these transitions can be deduced from the peak positions in the spectral function.
Denoting these energies with $\Delta \epsilon_{Li}$, $\Delta \epsilon_{iR}$ and $\Delta_{ij}$
we can identify, at HF level, transitions at energies $\Delta \epsilon_{L3,2R} \approx 0.5$, $\Delta \epsilon_{L4} \approx 0.9$,
$\Delta_{23} \approx 1.6$, $\Delta_{34} \approx 1.3$, $\Delta \epsilon_{L2,3R} \approx 2.25$ and $\Delta \epsilon_{L1,4R} \approx 3.3$.
The largest peaks are due to the HOMO-LUMO transition $\Delta_{23}$ and the transitions  
$\Delta \epsilon_{L4}$, $\Delta \epsilon_{L3,2R}$ and we see that these are the only peaks that 
are clearly visible for the 2B/GW case,
since in the correlated case the transients and dipole moments are strongly damped.\\
\\
We now address the case of the AC block bias $U_1 (t)$.
In Fig. \ref{fig4} we show the corresponding dipole moments and transient
currents.
As in the DC bias case 
we observe that the GW results are very close to the 2B results. The currents and dipole moments show a similar initial 
transient behavior as in Fig. \ref{fig3} up to $t=20$ after which the bias switches sign. 
The inset in the Fig. \ref{fig4} a) contains the Fourier transform of the dipole moment
which displays peaks at the odd harmonics $(2n+1)\omega_0$ of the basic driving frequency $\omega_0$ of the AC bias.
The appearance of odd harmonics is a consequence of the even inversion symmetry of the unbiased Hamiltonian and  the odd inversion symmetry of the applied bias.  
We see up to 9 harmonic peaks which indicates  
that we are far from the linear response regime. 
From the inset of Fig. \ref{fig4} we also see that the harmonics at energies $1.4$ (harmonic order 9) and $1.7$ 
(harmonic order 11) are enhanced. This is due to strong mixing with the HOMO-LUMO transition that occurs at energy 1.6 in the DC case
(see fig.\ref{fig3}).

High harmonics with even order will develop when we
break the symmetry of the unbiased Hamiltonian or of the applied bias.
This happens for the case of the on-off AC bias voltage $U_2 (t)$ (see Eq.(\ref{eq:onoff})). The dipole moments and transients for this
case are displayed in Fig. \ref{fig5}.
The applied bias in this case has either even or odd inversion symmetry depending  on the time interval considered. Therefore even harmonics at frequencies
 $2n \omega_0$ are also visible 
in the Fourier transform of the dipole moment (Fig. \ref{fig5} inset). Also in this case the higher harmonics mix strongly with the main 
electronic transition frequencies leading to an increased intensity of the harmonics around the HOMO-LUMO gap frequency $\omega\approx 1.5$.
Furthermore there are increased harmonic peaks at frequencies $\omega \approx 0.5, 0.9$ which are due to mixing of the harmonics with
the transitions $\Delta \epsilon_{L3,2R}$ and $\Delta \epsilon_{L4}$ (see fig.\ref{fig3}).

\begin{figure}[t]
\centering
\includegraphics[width=1.0\textwidth]{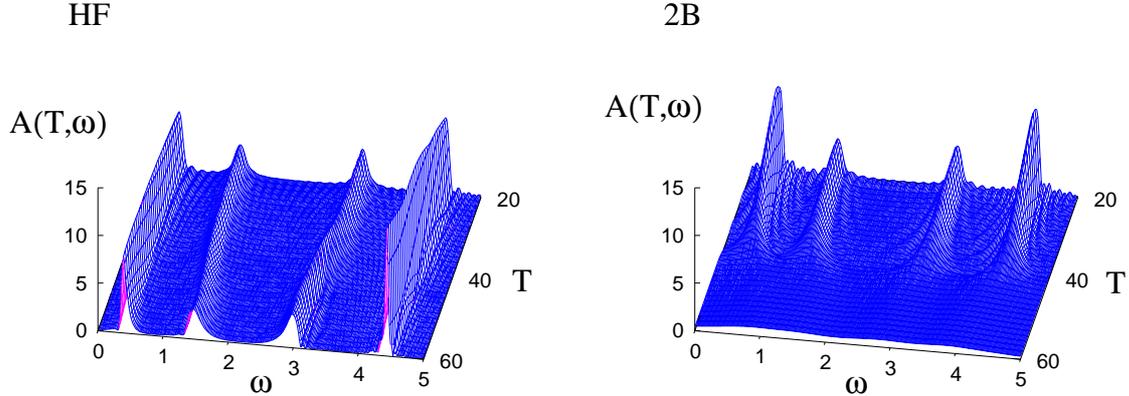}
\caption[]{Time-dependent spectral function $A(T,\omega)$ in the HF (left panel) 
and 2B (right panel) approximation. The system parameters are the same as 
for Fig. \ref{fig5}.}
\label{fig6}
\end{figure}

Now we turn our attention to the spectral functions.
In Fig. \ref{fig6} we show the time-dependent spectral function $A(T,\omega)$ (per spin and shifted with the chemical potential $\mu$) for the HF and 2B approximations
for the case of the on-off bias $U_2 (t)$.
The system is first propagated
without the external bias up to 40 a.u. after which the AC bias is turned on.
We see that the spectral peaks in Hartree-Fock approximation remain sharp during the whole time-propagation and 
due to the applied bias a closing of the HOMO-LUMO gap takes place. 
In contrast to the HF approximation, the spectral peaks in 2B start to broaden and loose intensity
rapidly after the bias has been switched on and is
caused by enhanced quasiparticle scattering at finite bias.
\begin{figure}[t]
\centering
\includegraphics[trim = 20mm 00mm 00mm 10mm, clip, width=12cm]{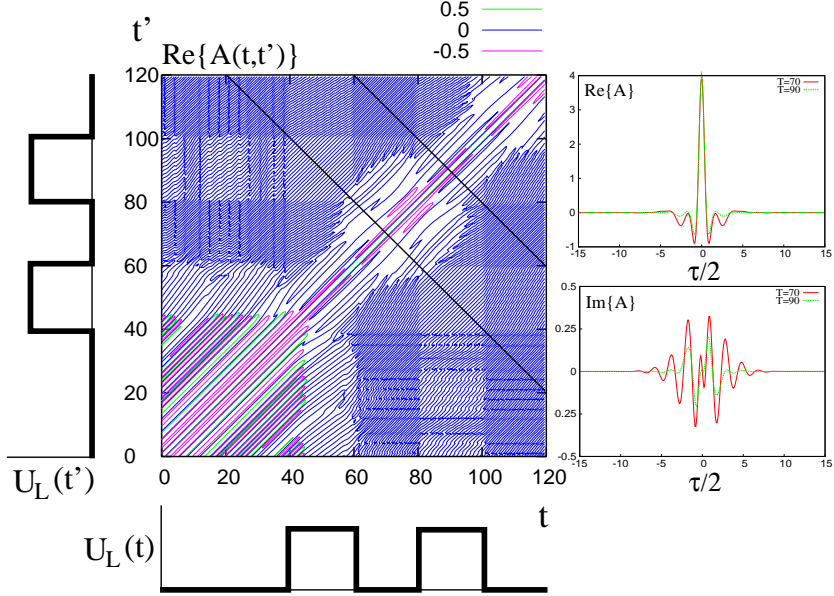}
\caption[]{Left panel: Real-time spectral function $A(t,t')$ in the 2B approximation for on-off AC bias voltage. 
Right panel: The real and imaginary parts of $A(T+\tau/2,T-\tau/2)$ as a function of $\tau/2$ for T=70 (solid line) and T=90 (dashed line). The corresponding 
cross-sections are indicated with black lines.}
\label{fig7}
\end{figure}
\begin{figure}[]
\centering
\includegraphics[trim = 20mm 00mm 00mm 10mm, clip, width=12cm]{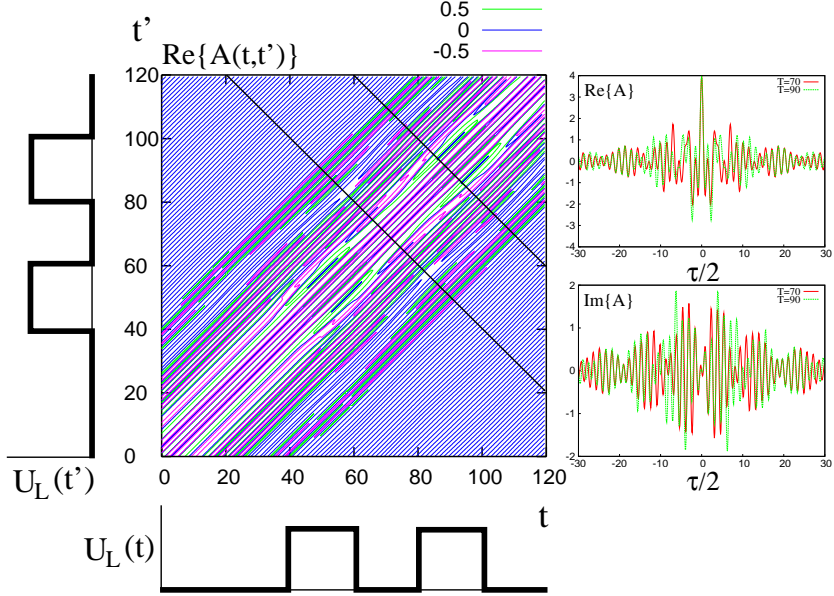}
\caption[]{Left panel: Real-time spectral function $A(t,t')$ in the HF approximation for on-off AC bias voltage . 
Right panel: The real and imaginary parts of $A(T+\tau/2,T-\tau/2)$ as a function of $\tau/2$ for T=70 (solid line) and T=90 (dashed line). The 
corresponding cross-sections are indicated with black
lines.}
\label{fig8}
\end{figure}
We note that we do not observe any noticeable shift in the position of the spectral peaks as a function of
the alternating bias. This can be explained from an analysis of the spectral function $A(t,t')$ in real time.
We display contour plots of these functions
in Figs. \ref{fig7} and \ref{fig8}. In these figures we also display
two center-of-time cross sections $A(T+\tau/2,T-\tau/2)$ for $T=70$ and $T=90$. Also for this case
the ground state was first propagated without a bias up to 40 a.u. after which the AC field was switched on.
Therefore for $t,t' \leq 40$ the contour plot describes the equilibrium system. For such times the spectral function
only depends on the relative time $t-t'$. For $t,t' >40$ the bias is applied and the spectral function starts to depend
on both time-coordinates separately. This is especially visible for the 2B case in Fig.\ref{fig7} where for
$40 < t,t' \lesssim 50$ we see a sudden collapse of the spectral function (we have checked this also for other values of the contour lines than displayed in the plot). 
This time interval coincides with the transient time of the current as shown
in Fig.\ref{fig5}. Note, however, that the spectral function $A(T,\omega)$ for average times $20 < T \lesssim 30$ is still dominated
by the equilibrium state since the Fourier transform over the relative time $\tau$ extends mainly over
values of $t,t'$ within the equilibrium region of the double time plane. For this reason the collapse of
the spectral function $A(T,\omega)$ of Fig.\ref{fig6} occurs within the time interval from $T=20$ to $T=40$. 
In Fig. \ref{fig7} we further observe a characteristic checkerboard pattern of the contour line at zero. This reflects
the various combinations (on-on, on-off and off-off) of the functions  $U_2(t)$ and $U_2(t')$ at times $t$
and $t'$. If we calculate the spectral function $A(T,\omega)$ by Fourier transforming over the relative time
coordinate $\tau$ we make an average over these regions.  This averaging makes the spectral function
less sensitive to the period of the applied AC bias. Nevertheless, some periodic changes in the cross
sections $A(T+\tau/2,T-\tau/2)$ can be observed. In the upper and lower right panels of Fig.\ref{fig7} we display
these cross sections at average times $T=70$ (center of the off-period) and $90$ (center of the on-period)
for both the real and the imaginary part of $A(T+\tau/2,T-\tau/2)$ as a function of $\tau/2$.  
On the time diagonal the real part of $A$ is always equal to $4$ since we have four states per spin in our
atomic chain.  Since the spectral function decays fast as a function of $\tau$
the main changes in the spectral function occur close to the time diagonal. When we Fourier transform these
cross section small shifts in the spectral peaks are observed. In the on-state at $T=90$ we found a slight closing
of the HOMO-LUMO gap as compared to the off-state at $T=70$.\\
In Fig.\ref{fig8} we display the spectral function $A(t,t')$ for the HF approximation.
It can be seen that the oscillations along the relative time coordinate $\tau$ in the spectral functions 
are much less damped in HF (Fig. \ref{fig8}) than in 2B (Fig. \ref{fig7}).
As a consequence we do not see the checkerboard pattern observed in the 2B case since it occurs much farther away from
the time diagonal. The functions $A(T+\tau/2,T-\tau/2)$ clearly show a slow and a fast oscillation as a function of $\tau/2$ corresponding to the two
outer and inner peaks (at positive and negative frequencies) of the time-dependent spectral function $A(T,\omega)$ displayed in Fig. \ref{fig6}.
Note that in this figure we shifted the peaks with the chemical potential $\mu=2.26$.\\

\begin{figure}[t]
\centering
\includegraphics[width=1.0\textwidth]{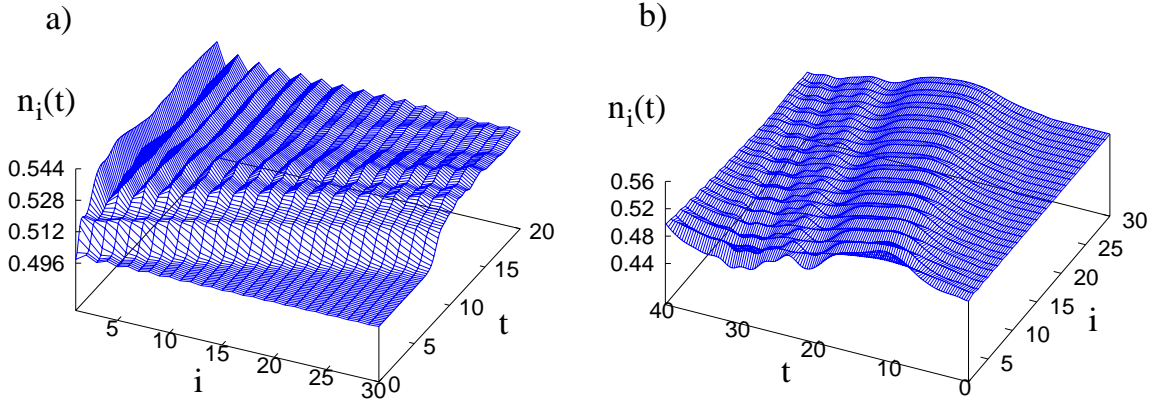}
\caption[]{Site occupation number $n_i (t)$ per spin in the right lead as a function of time $t$ and site label $i$.
The results are displayed
 for (a) DC bias, and (b) AC bias $U_1 (t)$ within the HF approximation. }
\label{fig9}
\end{figure}

We now turn our attention to the lead densities for the case of the AC block bias $U_1(t)$.
In right/left panel of Fig. \ref{fig9} we show the time evolution of the site occupation numbers $n_i (t)$ 
in the right lead for the AC/DC biases. These are calculated for up to 30 sites deep into the lead from Eq.(\ref{leaddens}) using $n_i (t)= -i \bcalG_{\a\a,rr}(t_{-},t_{+})$
where $r=i\sigma$.
For the case of a DC bias shown in panel (a) we
can clearly observe a density wave front that propagates into the right lead.
Furthermore the electron density 
displays a zigzag-pattern with an amplitude that  decreases away from the first site into the lead. These are the usual 
Friedel oscillations caused by the presence of the atomic chain in the lattice.
Their spatial profile is given by
$\delta n(x) \sim \sin(2k_Fx)$ where $k_F$ is the Fermi wave vector and $x$ the distance to the impurity site. 
In the case of the half-filled lead energy band that we consider 
$k_F=\pi/2$ corresponding to a density variation that alternates on every site. 
 When we switch on the bias in the leads the Friedel oscillations get enhanced by roughly an order
 of magnitude.
 
In panel (b) of Fig.\ref{fig9} we shown the lead density profile for the case of the AC bias $U_1(t)$.
As in the DC case, we see a wavefront moving into the lead. However, due to the periodic alternation of
the bias these wavefronts get superimposed on the returning ones giving rise to interferences
causing additional wiggles in the density profile.

\section{Conclusions}
\label{conc}


We proposed a time-dependent many-body approach based on the real time 
propagation of the KB equations for open and inhomogeneous systems. 

The method was applied to study transport dynamics through 
an atomic chain under AC and DC bias voltages. We calculated the 
spectral functions and the
time-dependent current, density and dipole moment within the 
HF, second Born and GW conserving approximations. 
We found that electron correlations beyond mean field have a large 
impact on time-dependent and steady state properties.
Both in the AC and DC case the HF spectral function retains its 
sharp structures upon applying a bias,
while the spectral function calculated within the 2B and GW
approximations broadens considerably.

The strong AC biases that we applied lead to highly nonlinear
perturbations. As a consequence
high-order harmonics of the driving frequency are observed in 
the time-dependent dipole moment.
Depending on the
symmetry of the Hamiltonian and of the applied AC voltage  
the odd and even harmonics are generated. The harmonics with largest 
intensity are those with energy close to the HOMO-LUMO 
gap.


\section*{References}

\begin{thebibliography}{9}

\bibitem{H2nature}
R. H. M. Smit \emph{et al}, 
Nature 419, 906-909 (2002).

\bibitem{science}
M. A. Reed \emph{et al}, 
Science Vol. 278. no. 5336, 252-254 (1997).

\bibitem{kjvhf.1991}
L. P. Kouwenhoven, A. T. Johnson, N. C. van der Vaart, C. J. P.
M. Harmans, and C. T. Foxon, 
Phys. Rev. Lett. {\bf 67}, 1626 (1991).

\bibitem{smcg.1999}
M. Switkes, C. M. Marcus, K. Campman, and A. C. Gossard,
Science {\bf 283}, 1905 (1999).

\bibitem{lbtsajwc.2005}
P. J. Leek, M. R. Buitelaar, V. I. Talyanskii, C. G. Smith, D.
Anderson, G. A. C. Jones, J. Wei, and D. H. Cobden, 
Phys. Rev. Lett. {\bf 95}, 256802 (2005).

\bibitem{zgsgm.2008}
Z.Zhong, N.M.Gabor, J.E.Sharping, A.L.Gaeta, and P.L. McEuen
Nature Nanotechnology {\bf 3}, 201 (2008)

\bibitem{tcd.2002}
A.Tikhonov, R.D.Coalson, and Y.Dahnovsky
J.Chem.Phys. {\bf 117}, 567 (2002)

\bibitem{bsin.2004}
R.Baer, T.Seideman, S.Ilani, and D.Neuhauser
J.Chem.Phys. {\bf 120}, 3387 (2004)

\bibitem{mgm.2007}
V.Moldoveanu, V.Gudmundsson and A.Manolescu
Phys.Rev.B {\bf 76}, 165308 (2007)

\bibitem{mmg.2009}
V. Moldoveanu, A. Manolescu, V.Gudmundsson, 
cond-mat arXiv:0909.0815.

\bibitem{skrg.2008}
G. Stefanucci, S. Kurth, A. Rubio and E. K. U. Gross,
Phys. Rev. B {\bf 77}, 075339 (2008).

\bibitem{mssvl.2008}
P. My\"oh\"anen, A. Stan, G. Stefanucci and R. van Leeuwen,
Europhys. Lett. {\bf 84}, 67001 (2008).

\bibitem{mssvl.2009}
P. My\"oh\"anen, A. Stan, G. Stefanucci and R. van Leeuwen,
Phys. Rev. B {\bf 80}, 115107 (2009).

\bibitem{dsvl.2006}
N. E. Dahlen, A.Stan and R.van Leeuwen
J. Phys. Conf. Ser. {\bf 35}, 324 (2006).

\bibitem{dvl.2007}
N. E. Dahlen and R. van Leeuwen,
Phys. Rev. Lett. {\bf 98}, 153004 (2007).

\bibitem{vfva.2009}
M. Puig von Friessen, C. Verdozzi, and C.-O. Almbladh,
condmat arXiv:0905.2061.

\bibitem{bbvlds.2009}
K. Balzer, M. Bonitz, R. van Leeuwen, N. E. Dahlen and A. Stan,
Phys. Rev. B {\bf 79}, 245306 (2009).

\bibitem{book2}
L. P. Kadanoff and G. Baym, 
{\em Quantum Statistical Mechanics} (Benjamin, New York, 1962).

\bibitem{d.1984}
P. Danielewicz, 
Ann. Phys. (N.Y.) {\bf 152}, 239 (1984).

\bibitem{kb.2000}
N.-H. Kwong and M. Bonitz
Phys. Rev. Lett. {\bf 84}, 1768 (2000).

\bibitem{w.1991}
M. Wagner, 
Phys. Rev. B {\bf 44}, 6104 (1991).

\bibitem{sa.2004}
G. Stefanucci and C.-O. Almbladh,
Phys. Rev.B {\bf 69}, 195318 (2004).

\bibitem{TDDFTbook}
R. van Leeuwen, N. E. Dahlen, G. Stefanucci, C. O. Almbladh,
and U. von Barth, 
{\em Time-Dependent Density Functional Theory}
(Springer, New York, 2006); Lect. Notes Phys. {\bf 706}, 33 (2006).

\bibitem{sdvl.2009b}
A. Stan, N. E. Dahlen and R. van Leeuwen,
J. Chem. Phys. {\bf 130}, 224101 (2009).

\bibitem{tr.2008}
K. S. Thygesen and A. Rubio
Phys. Rev. B{\bf 77}, 115333 (2008).

\bibitem{t.2008}
K. S. Thygesen,
Phys. Rev. Lett. {\bf 100}, 166804 (2008).

\bibitem{b.1962}
G. Baym
Phys. Rev. {\bf 127}, 1391 (1962).

\bibitem{bk.1961}
G. Baym and L. P. Kadanoff, 
Phys. Rev. {\bf 124}, 287 (1961).

\bibitem{vbdvls.2005}
U. von Barth, N. E. Dahlen, R. van Leeuwen and G. Stefanucci,
Phys. Rev. B {\bf 72}, 235109 (2005).

\bibitem{sdvl.2009}
A. Stan, N. E. Dahlen and R. van Leeuwen,
J. Chem. Phys. {\bf 130}, 114105 (2009).

\bibitem{dvl.2005}
N. E. Dahlen and R. van Leeuwen,
J. Chem. Phys. {\bf 122}, 164102 (2005).

\bibitem{mw.1992}
Y. Meir and N. S. Wingreen,
Phys. Rev. Lett. {\bf 68}, 2512 (1992).

\bibitem{jwm.1994}
A.-P. Jauho, N. S. Wingreen and Y. Meir,
Phys. Rev. B {\bf 50}, 5528 (1994).








\end{thebibliography}
\end{document}